\documentclass[]{spie}  

\usepackage{amsmath,amsfonts,amssymb}
\usepackage{graphicx}
\usepackage{comment}
\usepackage{caption}
\usepackage{subcaption}
\usepackage{tabularx}
\usepackage[colorlinks=true, allcolors=blue]{hyperref}

\title{Light scrambling and focal ratio degradation of thin multimode fibers with different core geometries}

\author[a]{Man-Yin Leo Lee}
\author[a]{Zhiheng Lin}
\author[a]{Chit-Ho Hui}
\author[a]{Renbin Yan}
\author[a]{YiuHung Cheung}
\author[a]{Horace Tsz-Hong Hung}
\author[b]{Matthew A. Bershady}
\author[c]{Sabysachi Chattopadhyay}
\author[b]{Michael P. Smith}
\affil[a]{Department of Physics, The Chinese University of Hong Kong, Shatin, New Territories, Hong Kong SAR, China}
\affil[b]{Department of Astronomy, University of Wisconsin, 475 N. Charter St., Madison, WI 53706, USA}
\affil[c]{South African Astronomical Observatory, 1 Observatory Rd, Observatory, Cape Town, 7925, South Africa}

\authorinfo{Further author information: (Send correspondence to M.-Y.L.L.)\\M.-Y.L.L.: E-mail: myleolee@link.cuhk.edu.hk, Telephone: 852 9283 6693}

\begin{document} 
\maketitle

\begin{abstract}
The performance of fiber-fed astronomical spectrographs is highly influenced by the properties of fibers. The near-field and far-field scrambling characteristics have a profound impact on the line spread function (LSF) of the spectra. Focal ratio degradation (FRD) influences the output beam size, thereby affecting the throughput, as well as the size of the collimator and dispersion elements. While previous research has indicated that these properties depend on the shape of the fiber core and showed that non-circular core fibers can yield uniform near-field scrambling, the result remains inconclusive for far-field. In this study, we investigate the near-field and far-field scrambling properties, along with the FRD, of 50-micron core fibers with different core geometries. We find that in addition to excellent near-field scrambling, octagonal-core fibers can also produce more uniform far-field output when compared to circular-core fibers. They also have less FRD effect when being fed with a f/3 beam.
\end{abstract}


\section{INTRODUCTION}
\label{sec:intro}  
Optical fibers have been widely used in astronomical spectrographs due to their ability to guide and rearrange light from the focal plane of the telescope to the spectrograph slit. This enables a cost-effective way to construct multiple object spectrometers (MOS) and integral field spectrographs (IFS). Galaxy surveys like SDSS IV MaNGA \cite{bundy_overview_2015, yan_sdss-iv_2016} made use of such advantages to provide spatially resolved spectroscopic observations on more than 10000 galaxies. Such surveys have been crucial to the advancement of understanding in interstellar medium and galaxy evolution. In the coming years, new instruments like the Slit Mask IFU (SMI) for the Southern African Large Telescope (SALT) \cite{chattopadhyay_slit_2022} and the Affordable Multiple Aperture Spectroscopic Explorer (AMASE) \cite{yan_design_2024, yan_prototype_2020} will continue to exploit such technology to provide spatially resolved spectrum on nearby galaxies and local HII regions.

The performance of fiber-fed spectrographs is greatly affected by the scrambling properties and focal ratio degradation of the optical fibers. These properties have been shown to have strong dependencies on the core geometries of the fibers \cite{spronck_extreme_2012}. Near-field refers to the distribution of light on the fiber output surface. As the spectral lines are images of the fiber output, variations of near-field affect the shape of the lines directly. Far-field refers to the intensity distribution of the fiber output beam. Non-uniformity of the far-field distribution results in uneven illumination of the spectrograph optics and variation of LSF on the detector. The phenomenon where the incoming beam becomes faster after passing through the fiber is referred to as FRD. Such widening determines the size of the collimator and the grating. Therefore, it is necessary to characterize these properties when constructing fiber-fed spectrographs.

The AMASE spectrograph is a wide-field fiber-fed integral field spectrograph that aims to explore HII regions in the Milky Way and galaxies in the local group\cite{yan_design_2024, cheung_optical_2024}. With high spatial (0.1 - 100 pc) and spectral (R $\approx$ 15000) resolution, the AMASE spectrograph will enhance the current understanding of the stellar feedback mechanism and the interstellar medium\cite{yan_prototype_2020}. As a wide-field survey, AMASE will cover 300 deg$^2$ in two years, targeting a wide range of objects such as M31, M33, LMC, SMC and HII regions within the Milky Way. To achieve the targetted angular resolution, the AMASE ferrule contains 547 50-micron core fibers with 85-micron center-to-center spacing. Under these circumstances, the fiber scrambling properties become essential as they determine the simplicity and efficiency of reconstructing the line spread function on the detector. On the other hand, the size of the grating and collimator will be determined by the effect of FRD.

This paper is structured as follows: We first describe the sample fibers we are testing in section \ref{sec: fibers}. Section \ref{sec: fiber test stand} describes the design of our fiber test stand and the procedures of telecentricity alignment before the test. Section \ref{sec: test procedure} gives the scrambling and FRD test procedures as well as the test results. Section \ref{sec: discussion and conclusion} concludes our main findings.

\section{SAMPLE FIBERS}
\label{sec: fibers}
We tested a total of six fibers in this study. Three of them have an octagonal core and the remaining three have a circular core. All of them are CeramOptec Optran WFNS fibers 1 meter in length. The octagonal (circular) fibers have core/cladding/buffer diameters of 50/75/124(125) microns. All of them have a numerical aperture of 0.22 $\pm$ 0.02 and SMA connectors on both ends. All the fibers were polished to 0.5 microns prior to the test.

\section{FIBER TEST STAND}
\label{sec: fiber test stand}
\subsection{DESIGN OF THE FIBER TEST STAND}
\label{subsec: fiber test stand}
\begin{figure}
	\includegraphics[width=\columnwidth]{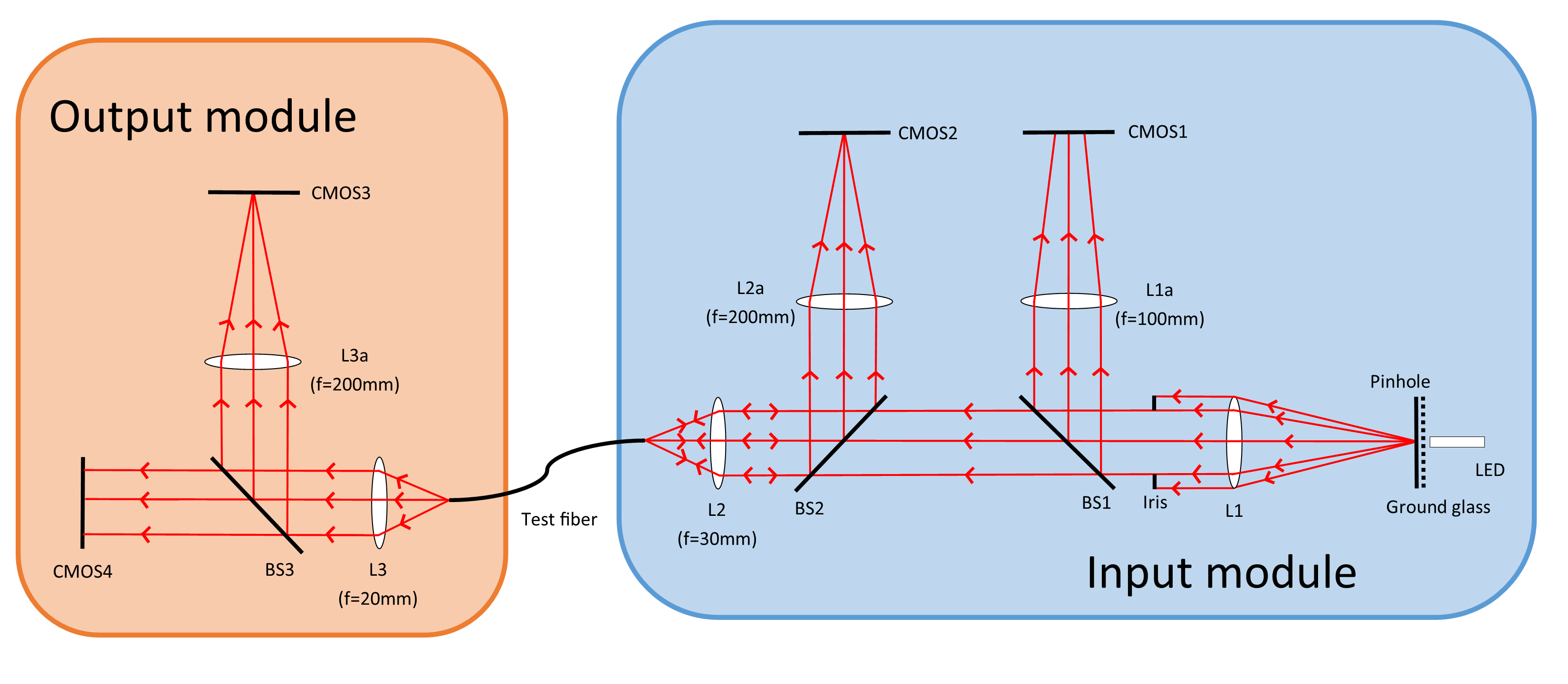}
    \caption{The schematic diagram of the fiber test stand.}
    \label{fig:fiber test stand}
\end{figure}

The fiber test stand was composed of two parts, the input module and the output module. The schematic view of the system is shown in Figure \ref{fig:fiber test stand}. 

A green LED centred at a wavelength of 520nm - 530nm was used as the light source. All dichroic beam splitters used in the experiments were Thorlabs CM1-BP108 Cube-Mounted Pellicle Beamsplitter. The mean transmittance and reflectance for unpolarized light between our light source-centred wavelength is 85.649\% and 14.218\%, respectively. CMOS1 and CMOS3 were two ZWO ASI120MM Mini. For CMOS2 and CMOS4, we used QHY533 and QHY163M respectively as larger detectors would be beneficial when conducting the tests.

The input module consisted of the light source, and two detectors (CMOS1 and CMOS2) to monitor the input light profile as well as the reflected image from the fiber surface. The light first passes through a opal diffuser and a pinhole that forms a spot as the input source. Pinholes with different sizes are used for different tests, as mentioned in the following sections. The beam is collimated by L1 before hitting the adjustable iris, which controls the beam size. The iris has an adjustable range of 2.5 to 10mm, which corresponds to an input focal ratio of f/12 to f/3. After passing through the iris, BS1 splits the beam to L1a and CMOS1. L1a is a 100mm doublet which allows the whole beam image to be captured by CMOS1, which acts as the monitor to the input flux. The system ensures the light source does not suffer from any obvious variations. 

The transmitted beam from BS1 reaches BS2. BS2 is placed in a configuration where the reflected beam will not be captured. The light is then focused by L2, which controls the demagnification of our object spot. Since EFFL$_{\text{L1}}$ is set to be 300mm and EFFL$_{\text{L2}}$ is set to be 30mm, the image size is then 1/10 of the original spot. Light reflected from the fiber surface will be captured by L2a and CMOS2 through reflection by BS2. We used CMOS2 to monitor the position of the input light spot on the fiber surface. Since the reflectance of the fiber surface is approximately 4\%, CMOS2 is only expected to have 4\%$\times$14.218\%=0.569\% of the total flux shining on the fiber surface neglecting any loss due to L2 and L2a.

The output module consists of two imaging systems monitoring the near-field and the far-field of the light coming out from the fiber. The light is first collimated by a microscopic objective L3 with an EFFL of 20mm and split by BS3. The reflected light from BS3 is captured by L3a such that a focused image of the fiber surface is captured by CMOS3, which corresponds to the near-field image. To better resolve the near field image, EFFL$_{L3a}$ is chosen to be 200mm, magnifying the output fiber surface 10 times before being captured by CMOS3. The transmitted light from BS3 is captured by CMOS4. This collimated light shows the distribution of light inside the light beam which is the far-field image.

\subsection{TELECENTRICITY ALIGNMENT}
\label{subsec: telecentricity}
\begin{figure}[h!]
    \centering
    \begin{subfigure}[b]{0.65\textwidth}
         \centering
         \includegraphics[width=\textwidth]{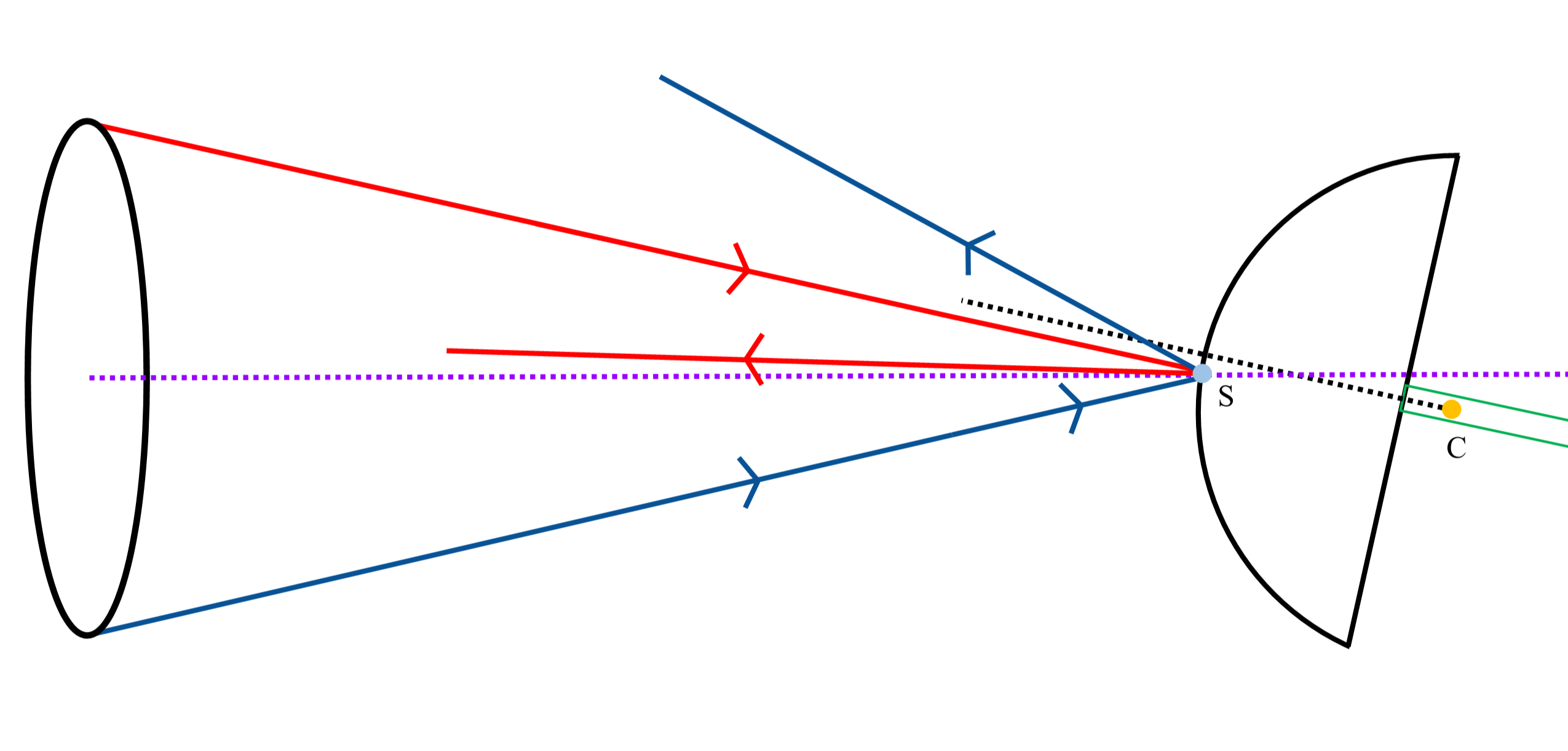}
         \caption{The light beam undergoes specular reflection on an arbitrary point S on the surface of the lens.}
         \label{subfig:telecentricity_surface}
    \end{subfigure}
    \begin{subfigure}[b]{0.65\textwidth}
         \centering
         \vspace{0.5cm}
         \includegraphics[width=\textwidth]{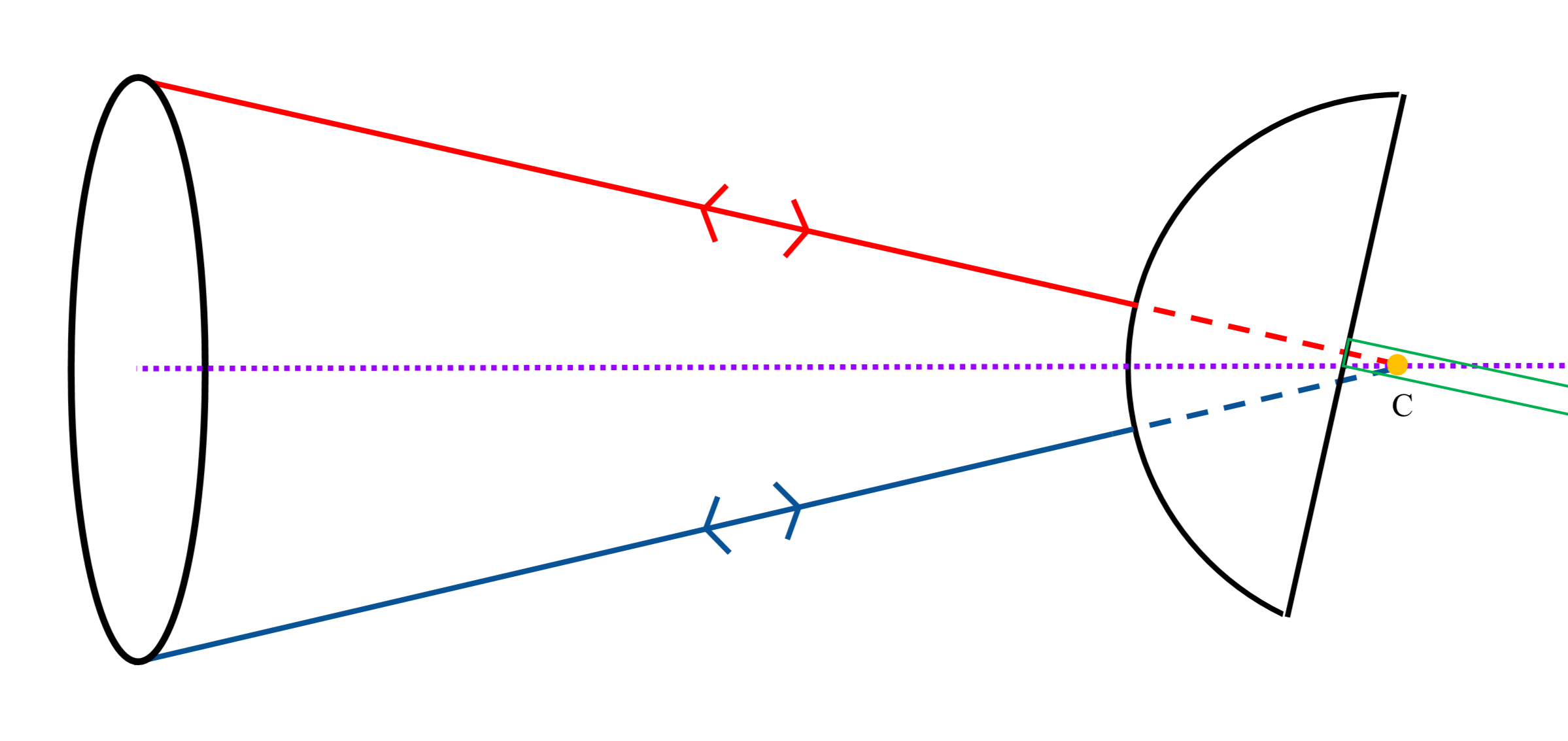}
         \caption{The light rays are reflected back following their incident path when passing through the center of curvature C of the spherical lens.}
         \label{subfig:telecentricity_center}
    \end{subfigure}
    \vspace{0.2cm}
	\caption{The telecentrcity alignment configuration. In the top panel, light rays are specularly reflected from the surface of the spherical lens. In the bottom panel, light rays are reflected back following its incident path. The system undergoes autocollimation when the images of the two scenarios overlap on the detector. C denotes the center of curvature of the spherical lens.}
    \label{fig:telecentricity}
\end{figure}
When testing different fibers, deviations in telecentricity of the input light are one of the factors affecting the their performances. To align the chief ray of the input converging light beam and the fiber surface, we performed an alignment process using a spherical lens for autocollimation. The spherical lens was placed in front of the fiber with the fiber located at the center of the lens's back surface. This lens-fiber system was then placed in front of the input module.

The distance between the surface of the lens and the input module was adjusted to the focal length of L2. Light converged by L2 underwent specular reflection on the lens surface, forming a dim spot on CMOS2. This is schematically shown in Fig. \ref{subfig:telecentricity_surface}. Since this spot is the image of specular reflection, any tilt or movement of the lens does not affect the position of the spot on CMOS2, as long as the beam is focused on the surface of the spherical lens. The position of this spot on the detector was recorded as a reference. The lens-fiber system was then moved to a position where the light rays pass through the center of curvature of the lens, as shown in Figure \ref{subfig:telecentricity_center}. Since the light rays pass through the center of curvature of the spherical lens, they were reflected back along the same path as the incidence light. A dim spot was then formed on CMOS2. To align the center of curvature of the spherical lens and the chief ray of the input beam, this dim spot was adjusted to the same position as the reference point set by the previous surface-reflected case. Next, the spherical lens was removed. We then adjusted the fiber tilt to compensate for the previous movements such that the input light spot overlapped with the fiber perfectly. After completing these procedures, the fiber tip and the chief ray should be aligned. Nevertheless, since the tilt in the last step was not completely along the fiber surface, which would also slightly affect the position of the fiber, this telecentric alignment procedure should be conducted iteratively to further ensure the telecentricity. The same procedure was applied to each fiber before testing. 

\section{TEST PROCEDURE AND DATA ANALYSIS}
\label{sec: test procedure}

\subsection{SCRAMBLING TEST}
\label{subsec: scrambling test}

\begin{figure}[h!]
	\includegraphics[width=\columnwidth]{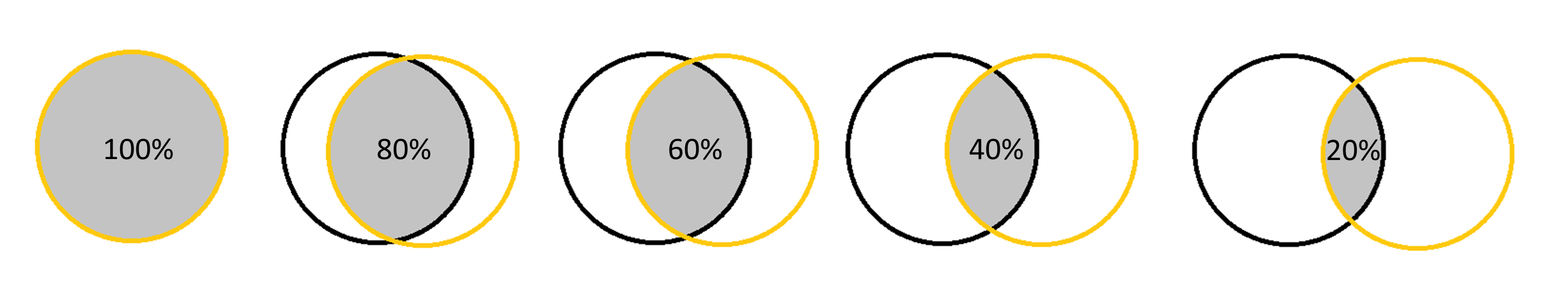}
    \caption{A demonstration of how we denote the overlapping cases. The yellow circle is the input light spot and the black circle is the fiber core. The distance to move from the perfectly centered case (100\%) is calculated by the overlapping area between the fiber core and the input light spot. The same distances were used on octagonal fibers. This figure is for illustration and the percentages of overlapping areas are not drawn accurately.}
    \label{fig:fiber coverage}
\end{figure}

To conduct the scrambling test, we used a 0.5mm pinhole behind the ground glass diffuser. This formed a 50-micron spot on the fiber surface, matching the core of the fibers and the input spot of the AMASE telescope after defocusing. The position of the spot was monitored by CMOS2 while the near-field and far-field profile of the fiber output was monitored by CMOS3 and CMOS4, respectively. The fiber input end was placed on a motorised stage that controls the fraction of the input light spot overlapping with the fiber core. Here, we denote the cases by the overlapping percentage between the spot and the fiber core, as shown in Figure \ref{fig:fiber coverage}. Hence, the perfectly centered case will be 100\%.

\begin{figure}[h!]
    \centering
    \begin{subfigure}[t]{\textwidth}
         \centering    
         \includegraphics[width=\textwidth]{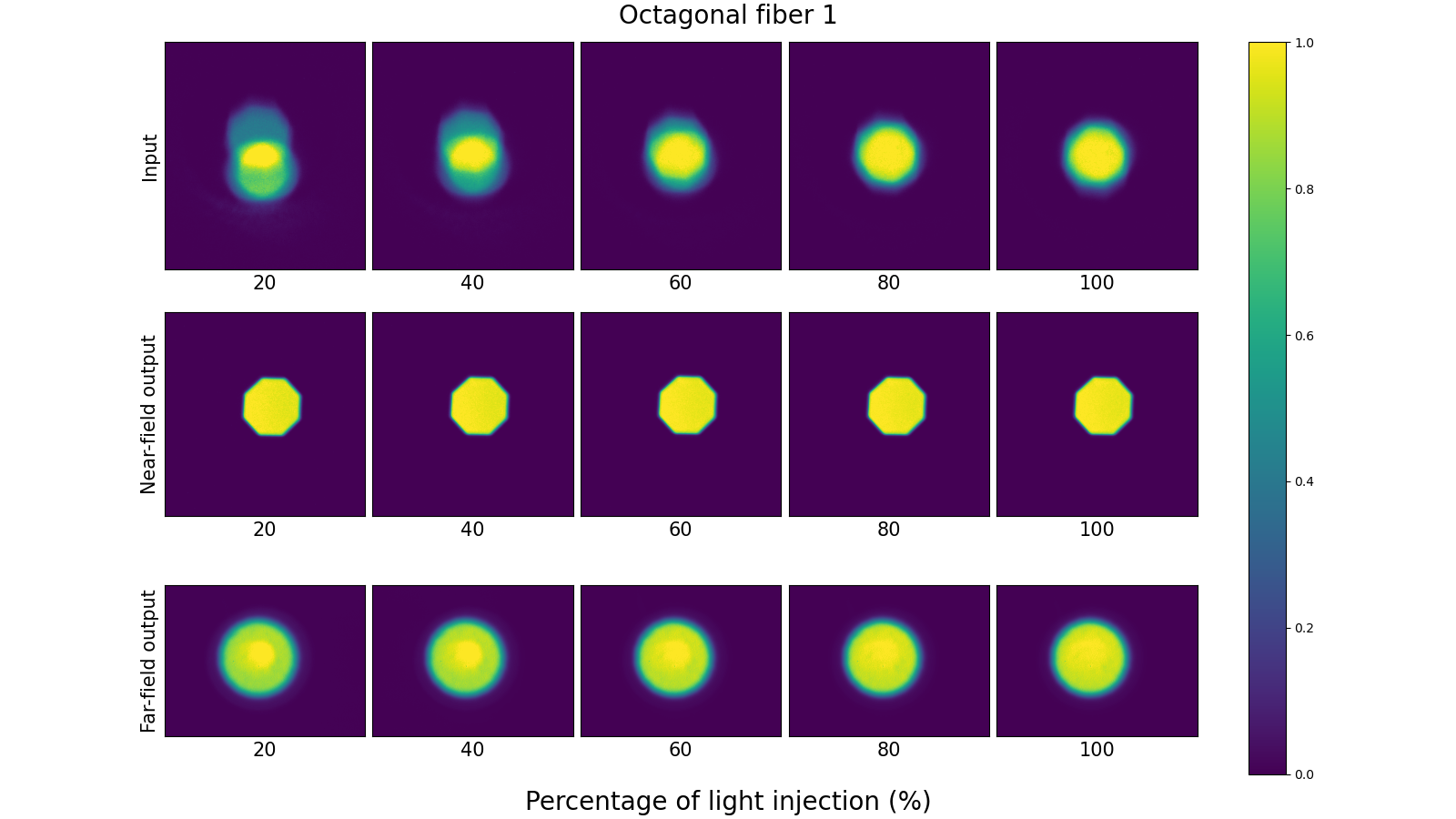}
         \caption{Scrambling test result images for one of the octagonal fibers.}
         \label{subfig:scrambling_image_oct}
    \end{subfigure}
        \begin{subfigure}[b]{\textwidth}
         \centering
         \vspace{0.5cm}
         \includegraphics[width=0.9\textwidth]{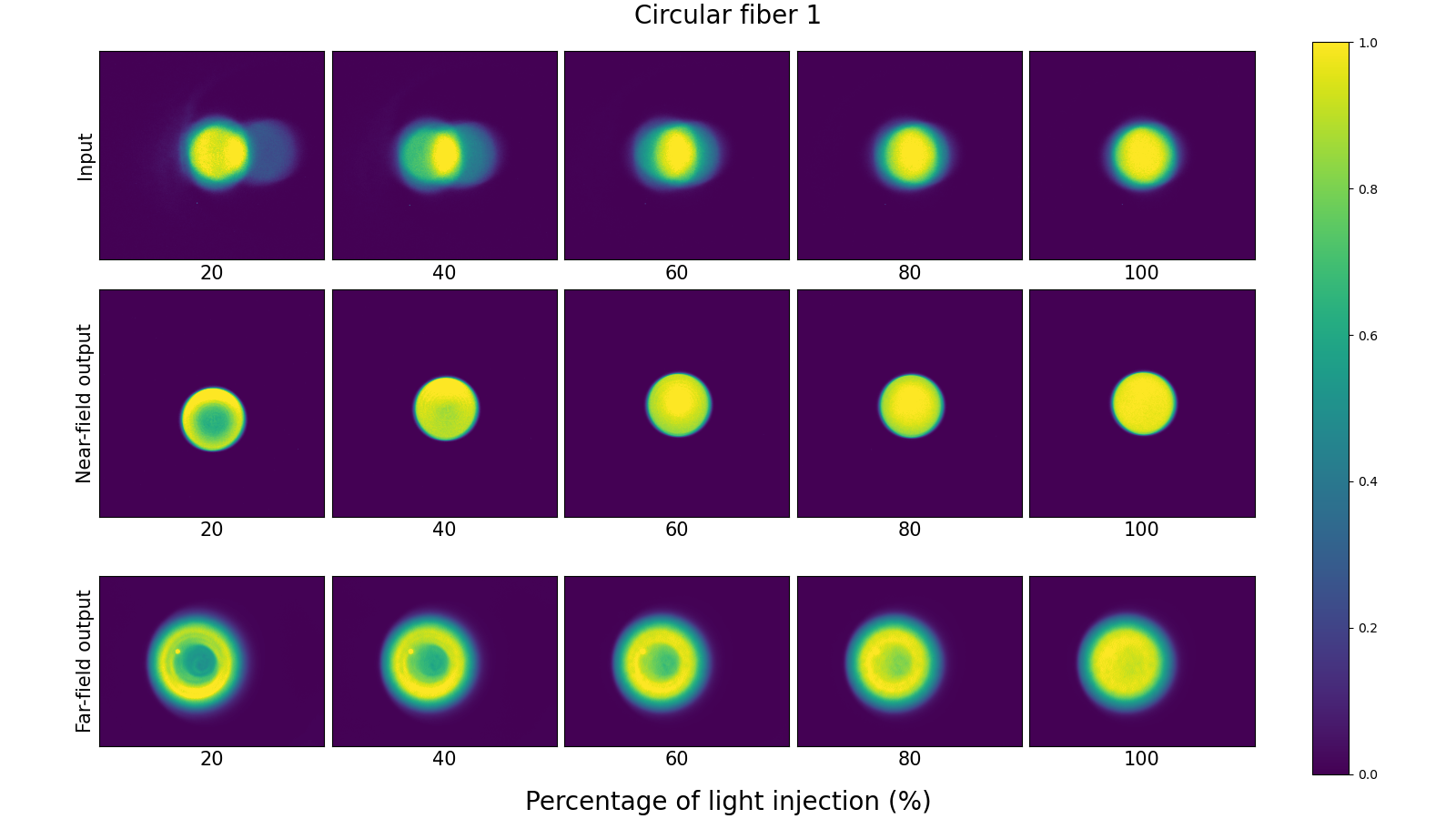}
         \caption{Scrambling test result images for one of the circular fibers.}
         \label{subfig:scrambling_image_circ}
    \end{subfigure}
    \vspace{0.2cm}
	\caption{The input, near-field and far-field output of two fibers having different core geometries under different light injection percentages.}
    \label{fig:scrambling_image}
\end{figure}

The scrambling test results for two of the fibers are shown in Figure \ref{fig:scrambling_image}. The near-field images for the octagonal fiber are very stable, as expected. On the other hand, with the light spot moving away from the fiber core, the near-field output of the circular fiber shows slight variation. The effect can be most obviously seen when the amount of light injection drops to 20\% of the perfectly centered case, where the flux in the center part of the fiber surface slightly drops. 

The far-field shows a similar result that the octagonal fiber tends to have a more stable intensity distribution. Although there are slight variations among the profiles, most of the pixels have flux values higher than 80\% of the brightest pixel. Similar to the near-field output, the circular fiber also shows center darkening feature when the percentage of light injection drops from 80\% to 60\%. Individual flux can drop below 70\% of the brightest pixel.

\begin{figure}[h!]
    \centering
        \includegraphics[width=0.98\columnwidth]{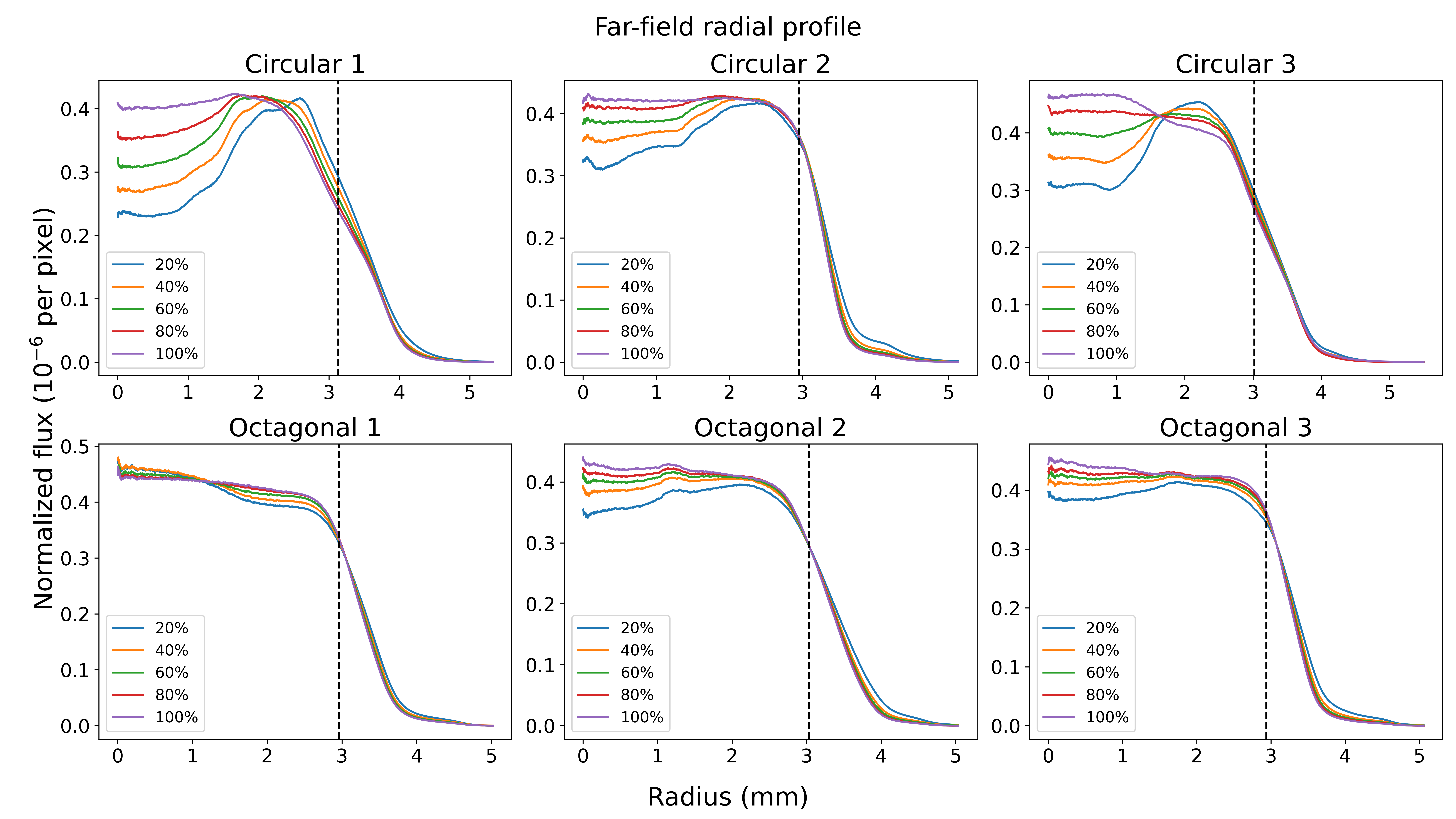}
    \caption{Azimuthally-averaged far-field radial profiles of the sample fibers. The flux is normalized by the total flux within the defined circle, and the black dashed line denotes EE80 of the 100\% profile. See subsection \ref{subsec: scrambling test} for details.}
    \label{fig:farfield_radial}
\end{figure}

\begin{figure}[h!]
	\includegraphics[width=0.98\columnwidth]{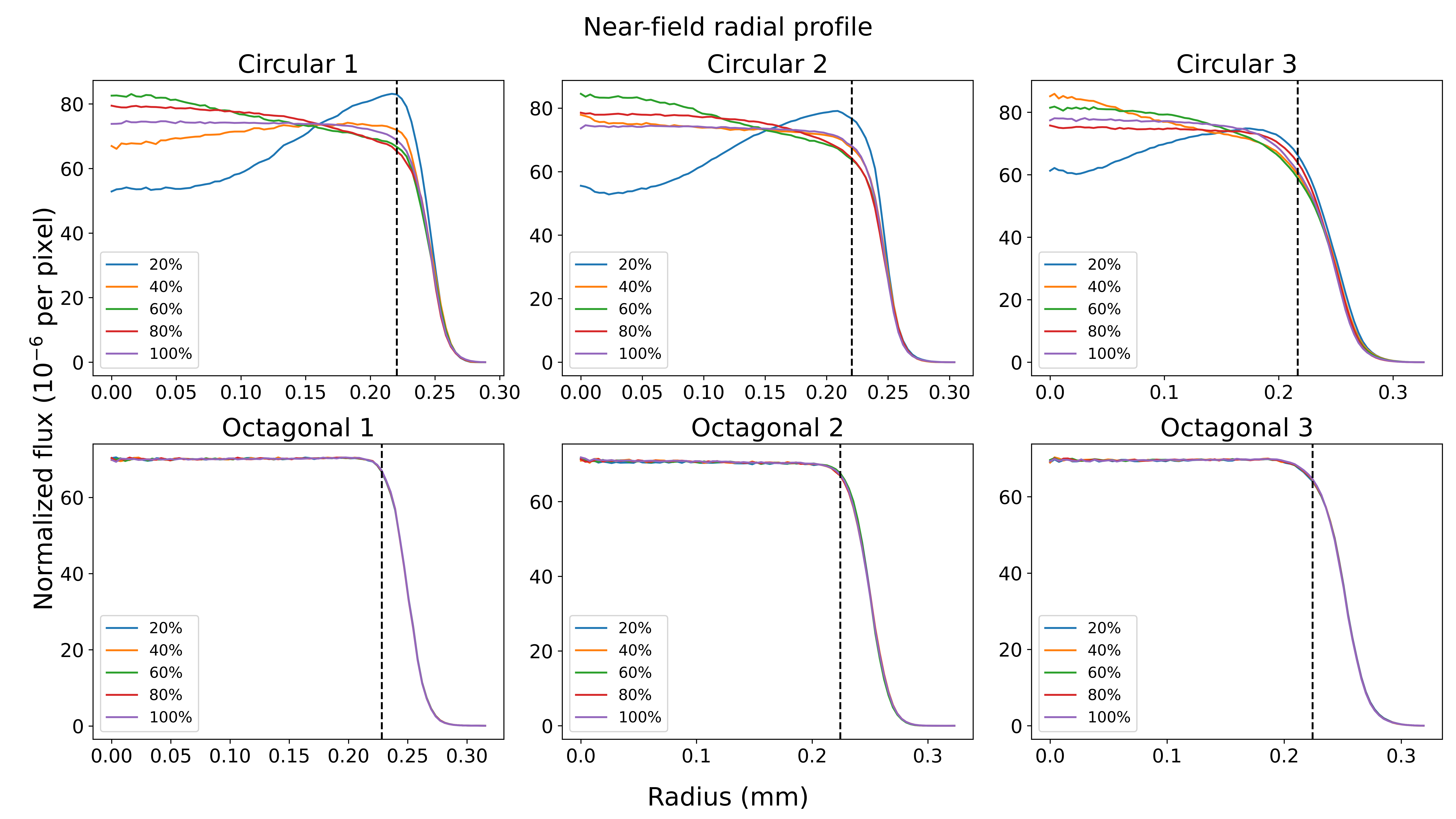}
    \caption{Azimuthally-averaged near-field radial profiles of the sample fibers. The flux is normalized by the total flux within the defined circle, and the black dashed line denotes EE80 of the 100\% profile. See subsection \ref{subsec: scrambling test} for details.}
    \label{fig:nearfield_radial}
\end{figure}

We further analyzed the data by plotting the azimuthally-averaged radial profile of each near-field and far-field outputs. A circle was fitted to the profile and the curve of growth of each image was plotted. The boundary of the circle was defined to be the distance where the curve of growth starts to drop. This is because negative flux should not present physically and such decrement in flux indicates a noise-dominated background region. Then, the azimuthally-averaged radial profile was computed. The flux of each pixel was normalized by the total flux within the circle. The radial profiles of the far-field and near-field images are shown in Figure \ref{fig:farfield_radial} and \ref{fig:nearfield_radial}, respectively. It is worth mentioning that the near-field image of the octagonal fiber was not a circle but this has a negligible impact on the azimuthally-averaged profile. From the figures, it is obvious that the octagonal fibers have more stable far-field and near-field profiles compared to their circular counterparts. Both the far-field and near-field profiles of circular fiber were affected by the percentage of light spot overlapping with the fiber core and suffered from a decrement of flux in the center when only one side of the fiber was illuminated. The same effect was seen in the far-field profiles of octagonal fibers, but their profile variations are more trivial.

We then computed the fractional deviation curve of each profile relative to the perfectly centered case. From the curve of growth, we define the radius of the circle that include a certain percentage of the total flux as `EE' (Encircled energy). For example, the radius of the circle which includes 80\% of the total flux will be EE80. Since we want to emphasize the bright part of the profiles, we only include pixels up to EE80 in the following analysis. We first computed the fraction deviation between the perfectly centered cases and each of the other cases. Then, for each cases, we calculated the root-mean-squared fractional deviation of each pixel. Results of the far-field and near-field profiles are shown in Table \ref{table: farfield_rms} and \ref{table: nearfield_rms}. As expected, the circular fibers have a higher fractional deviation than the octagonal fibers. For octagonal fibers, the far-field profile within EE80 will only change for about 10\% when the light spot change from perfectly aligned to the fiber center to the edge lightening 20\% of the fiber core. On the other hand, the near-field profile of octagonal fibers varies only up to 0.5\%, which is extremely stable. The near-field profile of the circular fibers varies around 5\% when 40\% and 60\% of the fiber are overlapping with the input light spot, but the number rises to more than 13\% when the light spot moves further away covering only 20\% of the fiber core. These test results concluded that octagonal fiber maintained a stabler scrambling profile both in far-field and near-field than the circular fibers.
\begin{table*}
\caption{Root-mean-squared value of the fractional deviation of far-field profiles from the 100\% case. Only the pixels up to EE80 of the 100\% case are considered. See subsection \ref{subsec: scrambling test} for details.}
\label{table: farfield_rms}
    \centering
    \begin{tabularx}{\linewidth}{@{}l *4{>{\centering\arraybackslash}X}@{}}
    \hline
       Fiber &   20\% light &   40\% light  &   60\% light &   80\% light\\
    \hline
      Circular 1 & 29.246\% & 21.336\% & 14.285\% & 7.105\% \\
      Circular 2 & 15.372\% &  9.713\% &  5.653\% & 2.07\%  \\
      Circular 3 & 23.123\% & 16.162\% &  9.867\% & 4.448\% \\
      Octagonal 1 &  4.606\% &  3.361\% &  1.62\%  & 0.504\% \\
      Octagonal 2 & 10.596\% &  5.636\% &  3.378\% & 1.76\%  \\
      Octagonal 3 &  8.942\% &  4.95\%  &  3.189\% & 1.854\% \\
    \hline
    \end{tabularx}
\end{table*}

\begin{table*}
\vspace{0.3cm}
\caption{Root-mean-squared value of the fractional deviation of near-field profiles from the 100\% case. Only the pixels up to EE80 of the 100\% case are considered. See subsection \ref{subsec: scrambling test} for details.}
\label{table: nearfield_rms}
    \centering
    \begin{tabularx}{\linewidth}{@{}l *4{>{\centering\arraybackslash}X}@{}}
    \hline
        Fiber &   20\% light &   40\% light &   60\% light &   80\% light\\
    \hline
    Circular 1 & 20.263\% & 5.404\% & 6.532\% & 4.685\% \\
    Circular 2 & 18.463\% & 1.448\% & 7.971\% & 4.086\% \\
    Circular 3 & 13.062\% & 5.109\% & 3.449\% & 2.822\% \\
    Octagonal 1 &  0.513\% & 0.325\% & 0.367\% & 0.493\% \\
    Octagonal 2 &  0.496\% & 0.606\% & 0.372\% & 0.315\% \\
    Octagonal 3 &  0.309\% & 0.315\% & 0.251\% & 0.413\% \\
    \hline
    \end{tabularx}
\end{table*}

\subsection{FOCAL RATIO DEGRADATION TEST}
\label{subsec: FRD test}
Focal ratio degradation (FRD) refers to the phenomenon that the focal ratio of the beam decreases after propagating through a fiber \cite{ramsey_focal_1988}. This increases optical etendue and requires the diameters of the collimator and grating to increase. It have been pointed out that the effect of FRD will be less distinctive when fed with a beam faster than f/4. In this study, we tested and compare the FRD performance of octagonal and circular fibers.

\begin{figure}[h!]
	\includegraphics[width=\columnwidth]{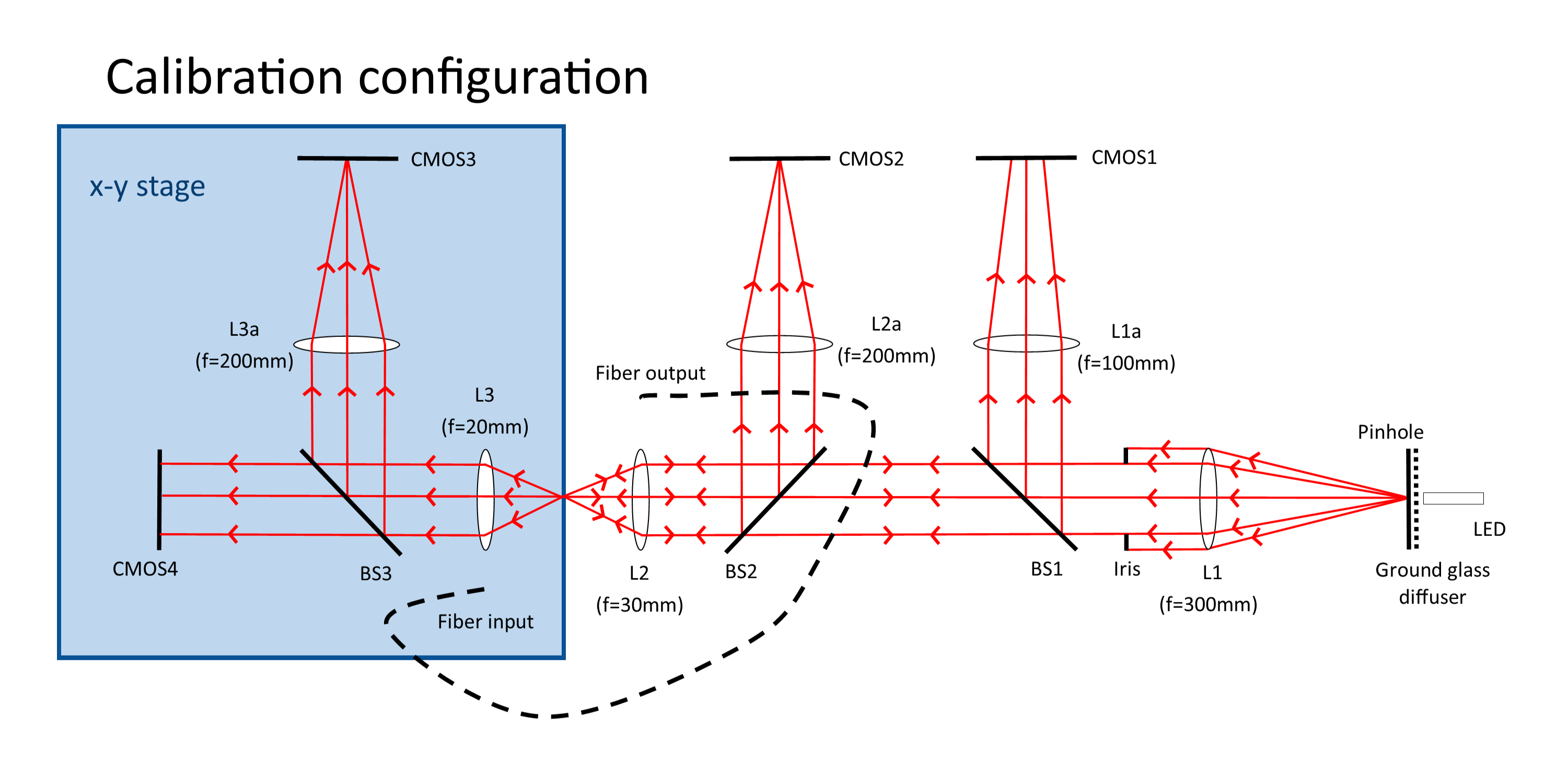}
    \caption{Schematic of the calibration configuration. The output module receives light directly from the input module and CMOS4 were used to provide the intensity distribution of the beam.}
    \label{fig:calibration mode}
\end{figure}

\begin{figure}[h!]
    \centering
    \includegraphics[width=\textwidth]{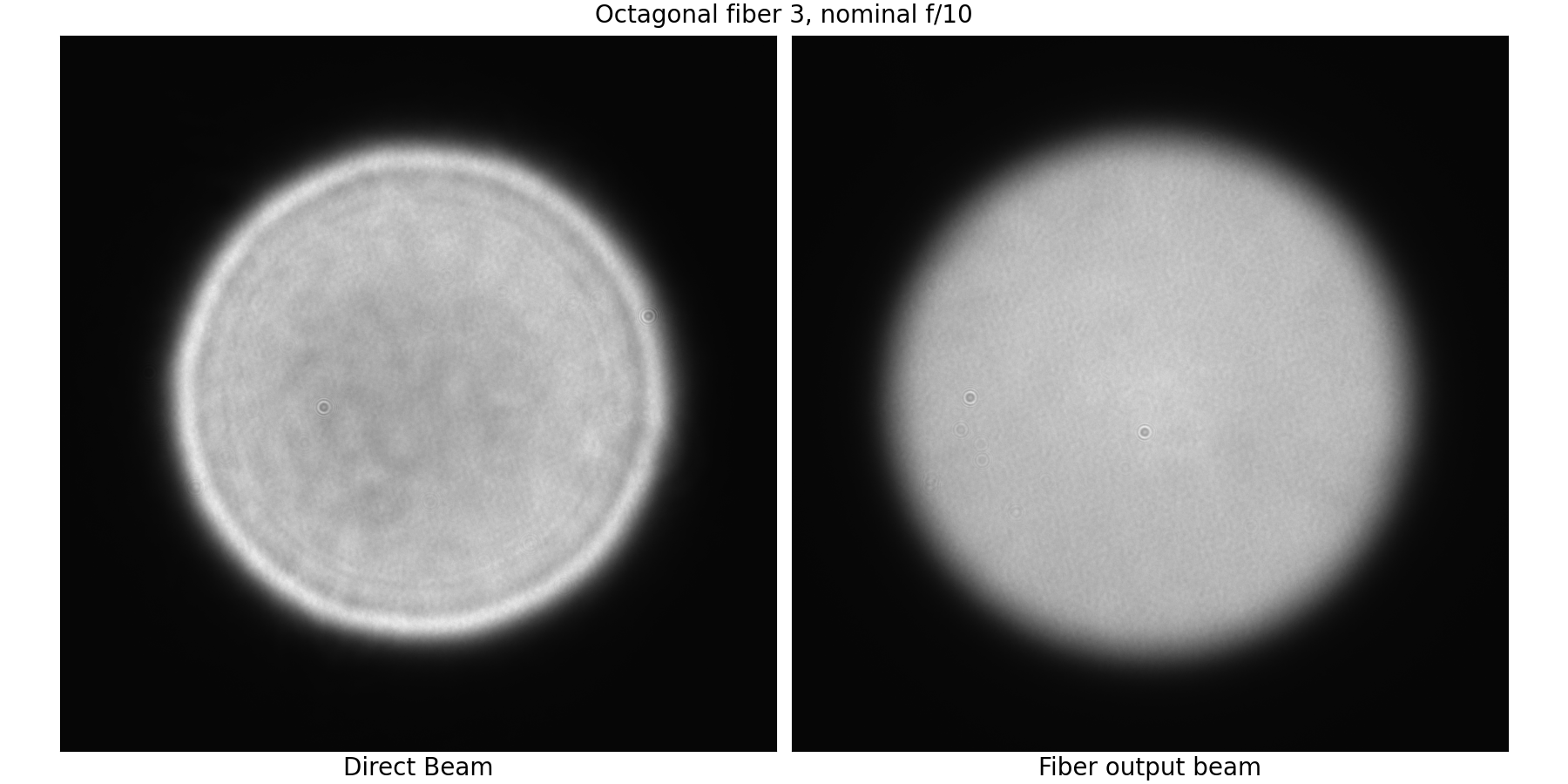}
    \vspace{0.1cm}
	\caption{Direct beam and fiber output beam of one of the octagonal fibers under the case of nominal f/3.}
    \label{fig:direct beam}
\end{figure}

We adopted the uncollimated beam approach \cite{carrasco_method_1994} to test the FRD performance of our sample fibers because it is differential and closer to the real application in astronomical spectrographs. To ensure the stability of flux injected into the fiber core, we used a 0.2mm pinhole, yielding a 20-micron light spot on the fiber surface in the experiment. The iris in Figure \ref{fig:fiber test stand} was adjusted to change the input f-number of the beam. Four input beam values, f/3, f/4, f/6 and f/12, were tested. However, since the accuracy of the iris cannot be guaranteed, we do not adopt this number as our input f-number. The f-number set by the iris will be referred to as the nominal f-number. After setting the iris size and doing the telecentricity alignment, the fiber test stand was converted into a calibration configuration in which the output module will receive light directly from the input module, as shown in Figure \ref{fig:calibration mode}. Under this configuration, CMOS4 were used to provide the intensity distribution of the direct beam. The switch in configuration was controlled by an x-y stage which interchanges the position of the fiber input with the output module. In the process, the tilt of the fiber was carefully conserved. After getting the intensity distribution of the direct beam, the test stand was switched back to the original configuration. Images from CMOS4, which represent the far-field output of the fiber, are captured. These procedures were repeated each time the iris size was adjusted. Figure \ref{fig:direct beam} shows the direct beam and fiber output beam of one of the fibers with nominal f/3.

\begin{figure}[h!]
    \centering
    \begin{subfigure}[b]{0.9\textwidth}
         \centering    
         \includegraphics[width=\textwidth]{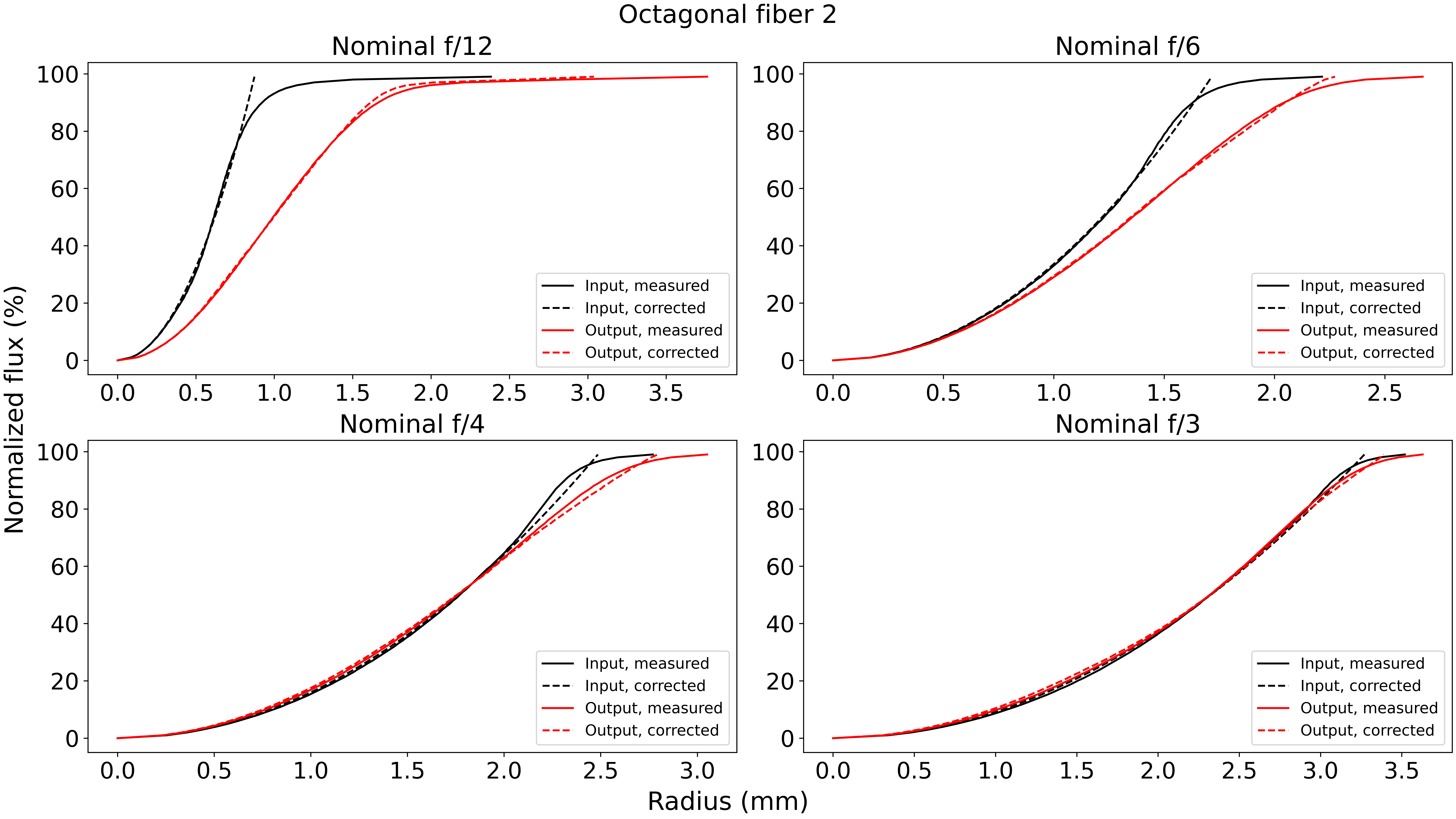}
         \caption{FRD measurements shown in the form of curves of growth for one of the octagonal fibers. }
         \label{subfig:curve of growth_oct}
    \end{subfigure}
        \begin{subfigure}[b]{0.9\textwidth}
         \centering
         \vspace{0.5cm}
         \includegraphics[width=\textwidth]{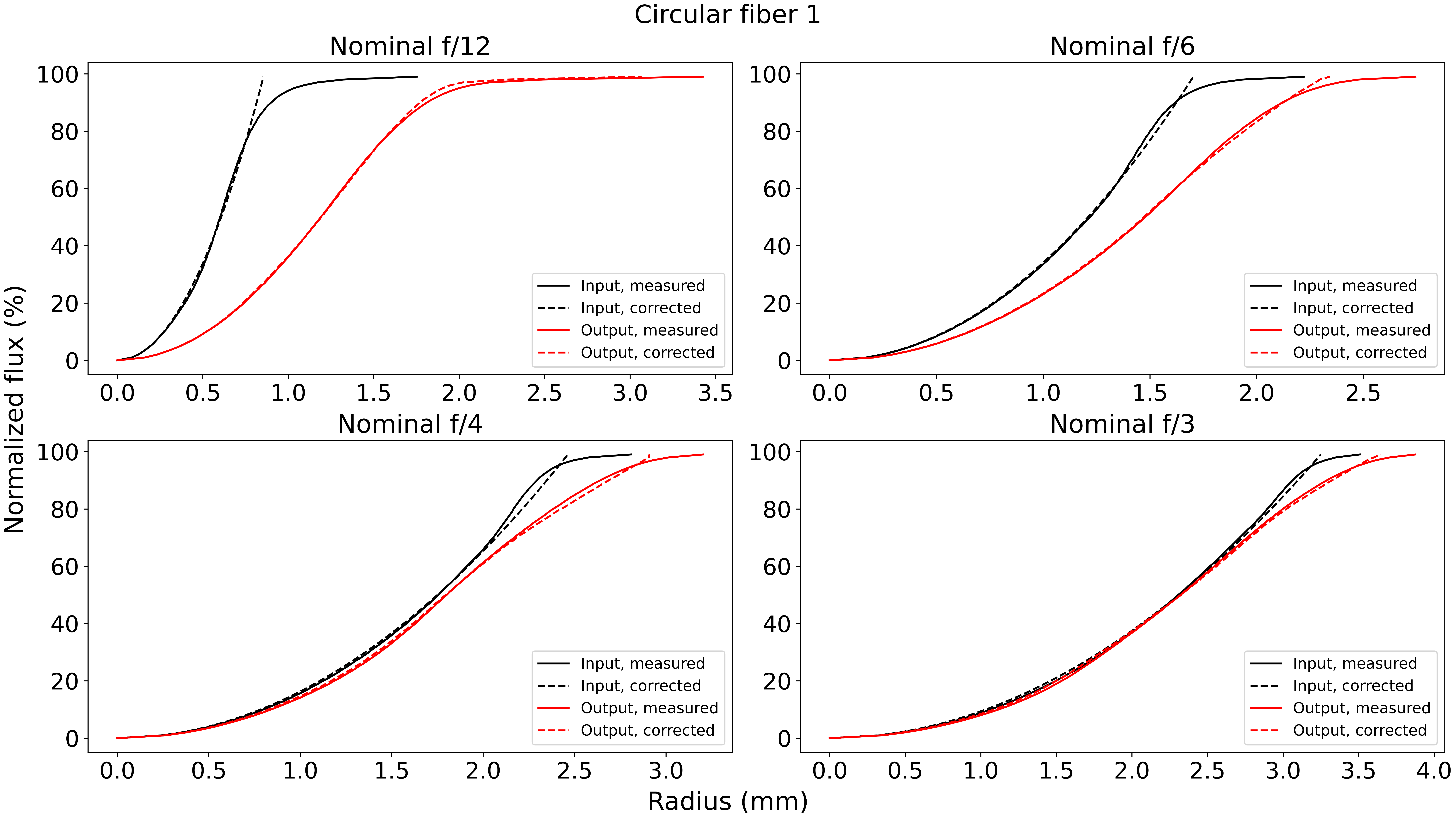}
         \caption{FRD measurements shown in the form of curves of growth for one of the circular fibers. }
         \label{subfig:curve of growth_circ}
    \end{subfigure}
    \vspace{0.2cm}
	\caption{FRD measurements shown in the form of curves of growth for two of the sample fibers under different nominal f-numbers. The correction was done as mentioned in subsection \ref{subsec: FRD test}. Note that for nominal f/3, the input and output beams are extremely closed for octagonal fibers, showing the effect of FRD is minimal when fed with such a fast beam.}
    \label{fig:curve of growth}
\end{figure}

We used the same method as the last subsection to determine the boundary of our direct beams. The derived beam diameter was defined to be the actual diameter and converted into the f-number input. Then, the curves of growth were computed and EE80, which is the radius of the circle encircling 80\% of total light, was defined. We then used the curve of growth up to EE80 to fit an ideal beam profile which had a constant cross-section flux distribution. The departure of the measured direct beam from the ideal beam was then used to correct for the output beam. The correction was done in quadrature to every radius at a given encircled energy level\cite{crause_investigation_2008}. Curves of growth and the correct beams for two sets of sample fibers are shown in Figure \ref{fig:curve of growth}. It is unexpected that in some of the cases, the direct beam has a smaller EE value than the ideal beam in the region of EE70 to EE90. This might be due to the diffraction pattern or edge brightening seen in the direct beam, as shown in Figure \ref{fig:direct beam}.

\begin{figure}[h!]
    \centering
    \includegraphics[width=0.75\textwidth]{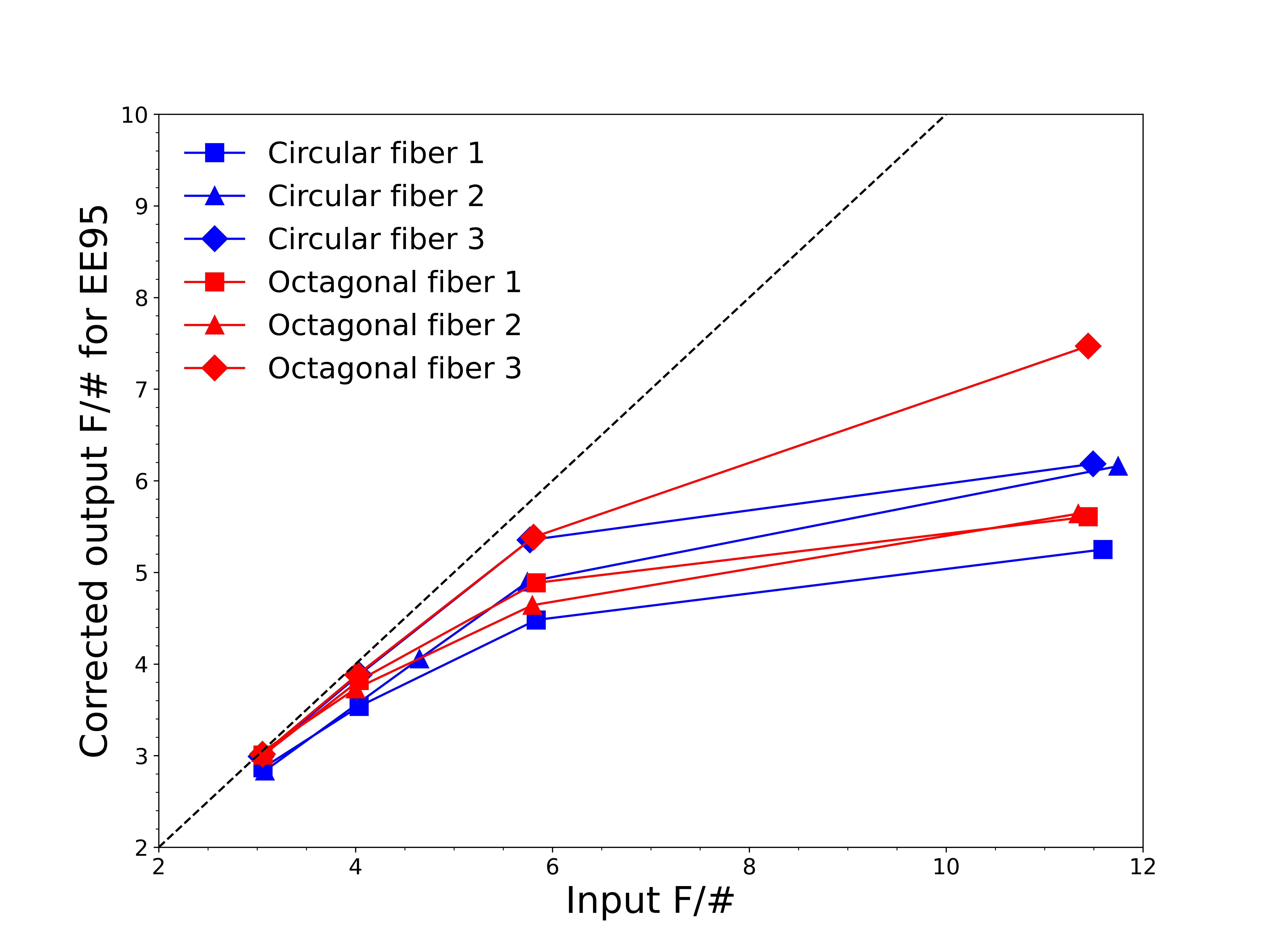}
	\caption{Output focal ratio derived using EE95 versus input focal ratio.}
    \label{fig:FF plot}
\end{figure}

The result of f-number output versus f-number input is shown in Figure \ref{fig:FF plot}. In the figure, the output focal ratio is defined by EE95 of the corrected output beam. In larger f-number cases, the FRD effects did not seem to correlate with the core geometry of the fibers. On the other hand, when the input beam is fast, the octagonal fibers have slightly better performance than the circular fibers. The corrected and uncorrected output f-number in the nominal f/3 cases are shown in Table \ref{table: f-number}. With similar input f-numbers, all the octagonal fibers have a higher output f-number than the circular fibers. The result indicates that octagonal fibers are more suitable to be used in astronomical spectrographs that have fast input beams because of their minimal FRD effect.

\begin{figure}[h!]
    \centering
    \includegraphics[width=0.75\textwidth]{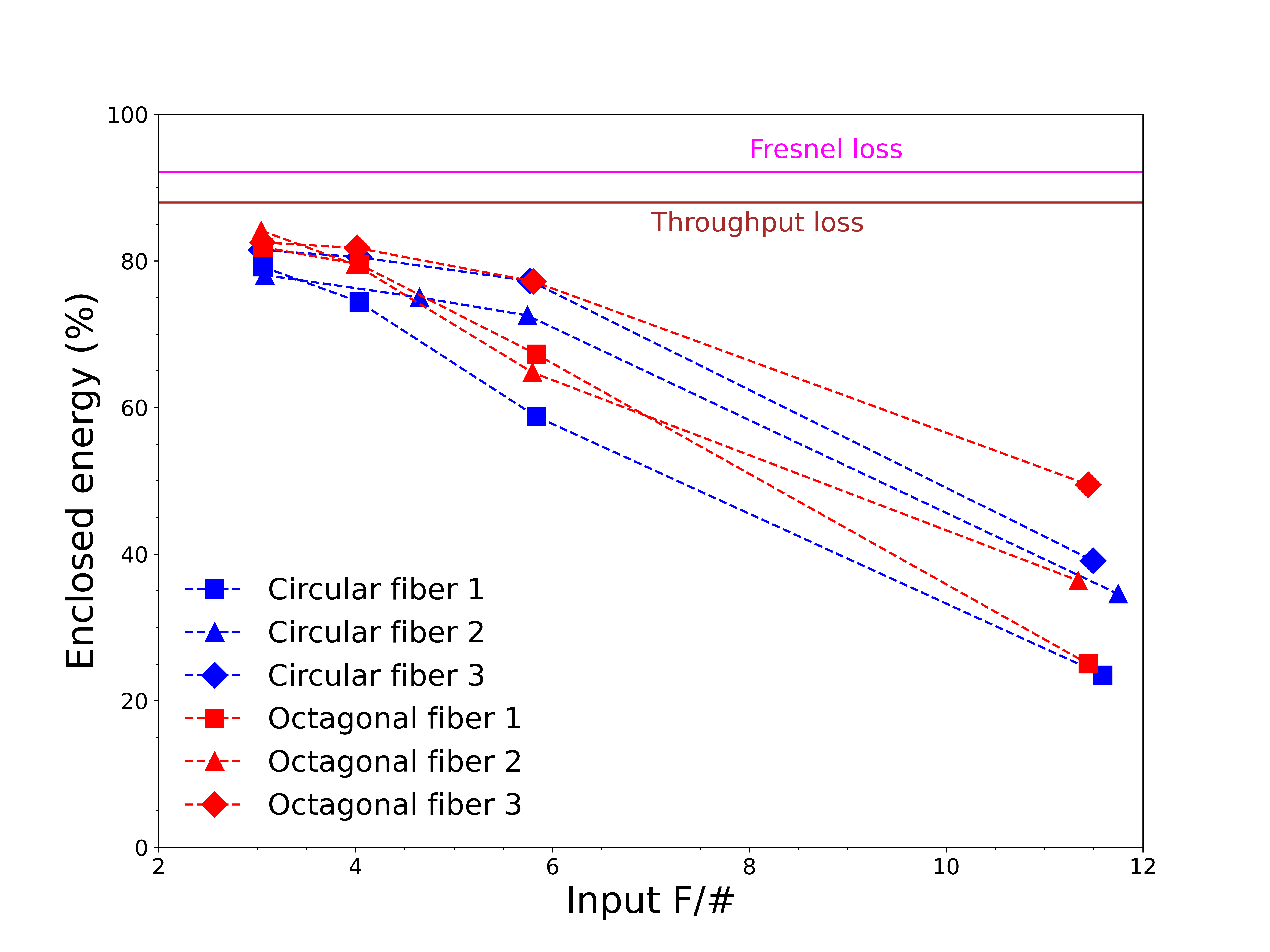}
	\caption{Enclosed energy in the output profile of a circle with the same radius of input beam versus input f-number. The magenta line indicates the expected Fresnel loss and the brown line represents the throughput loss.}
    \label{fig:EF plot}
\end{figure}

Figure \ref{fig:EF plot} shows the encircled energy in the output image using a circle with a radius same as the input beam. The encircled energies are plotted against the input f-number. Uncorrected beams are used when determining both the input and output energies. We labelled the expected Fresnel loss and the throughput loss. The Fresnel loss was calculated using the refractive indices of fused silica and represents the reflection loss on the fiber input and output faces. The throughput loss was the difference between the total energy in the direct beam and the output beam, both octagonal and circular fibers show similar throughput so we use the average value as the throughput loss. The energy loss due to FRD is the vertical distance between the label and the throughput loss line. The average FRD losses at nominal f/3 are 5.849\% and 9.555\% for octagonal and circular fibers, respectively.

\begin{table*}
\vspace{0.5cm}
\caption{The output f-number for the fibers under nominal f/3 input. The output f-number is defined by EE95 of the output beam. Both corrected and uncorrected results are presented.}
\label{table: f-number}
    \centering
    \begin{tabularx}{\linewidth}{@{}l *4{>{\centering\arraybackslash}X}@{}}
    \hline
        Fiber & Input f/\# &   Corrected output f/\# &   Uncorrected output f/\#\\
    \hline
  Circular 1 &       3.06  &                  2.867 &                    2.864 \\
  Circular 2 &       3.078 &                  2.834 &                    2.836 \\
  Circular 3 &       3.039 &                  2.99  &                    2.984 \\
  Octagonal 1 &       3.06  &                  3.007 &                    3.004 \\
  Octagonal 2 &       3.042 &                  3.035 &                    3.035 \\
  Octagonal 3 &       3.053 &                  3.016 &                    3.011 \\
    \hline
    \end{tabularx}
\end{table*}

\section{DISCUSSION AND CONCLUSION}
\label{sec: discussion and conclusion}
In this study, we tested the scrambling and focal ratio degradation performance on fibers with octagonal and circular cores. The fibers are 1-meter long, 50-micron core fibers with a numerical aperture of 0.22 manufactured by CeramOptec. We introduced our design of the fiber test stand and a unique way to ensure the telecentricity of the input beam to the fiber surface. The main results are as follows:

\begin{enumerate}
    \item Octagonal fibers have a very stable near-field scrambling profile, independent from the illumination region of the fiber input surface. The near-field profile of circular fiber changes with the position of the input light spot on the fiber surface.
    \item In terms of far-field profile, both octagonal and circular fibers show varying profiles when the input illumination region changes. Nevertheless, the profiles of the octagonal fibers are less affected.
    \item When being fed with a fast input beam like f/3, octagonal fibers have less FRD effect than their circular counterparts. In the case of f/3, octagonal fibers only lose 5.849\% of light due to FRD, whereas the circular fiber suffers from a 9.555\% loss.
\end{enumerate}

The test result indicates that, in modern astronomical spectrograph, like the ongoing AMASE spectrograph, octagonal fibers are more advantageous compared to circular fibers. Their near-field and far-field profiles are stable, hence providing a more uniform line spread function on the spectrograph detector. They are also less impacted by FRD when fed with a fast beam, which is most of the case in astronomical spectrographs. 

\acknowledgments 
We acknowledge support by the Hong Kong Jockey Club Charities Trust through the project, JC STEM Lab of Astronomical Instrumentation and Jockey Club Spectroscopy Survey System (2022-0052). We also acknowledge support by the CUHK Academic Equipment Grant and partial support by the grants from Research Grant Council of the Hong Kong Special Administrative Region, China [Projects No: 14302522 and 14303123].  RY acknowledges support by the Hong Kong Global STEM Scholar scheme. CH acknowledges support by the Hong Kong Innovation and Technology Hub through the Research Talent Hub program (GSP028). 

\bibliographystyle{spiebib} 

\end{document}